\documentclass[preprint,floatfix,noshowpacs,prb]{revtex4}
\usepackage{amsmath, amssymb}
\usepackage{graphicx}
\usepackage{times}
\usepackage{dcolumn}
\usepackage{array}
\usepackage{threeparttable}
\usepackage{setspace}
\emergencystretch=\maxdimen
\newcommand{\om}{\omega}
\newcommand{\be}{\begin{eqnarray}}
\newcommand{\ee}{\end{eqnarray}}
\newcommand{\fq}{\mathbf{q}}
\newcommand{\fk}{\mathbf{k}}
\begin{document} 

\title{Dynamic structure function of some singular Fermi-liquids}

\author{Chandra M. Varma}
\affiliation{Department of Physics, University of California, Riverside, CA. 92521}

\begin{abstract}
The density correlations of some singular Fermi liquids with anomalous properties such as resistivity varying linearly with T at low temperatures, a $T \log T$ contribution to the entropy and thermopower, etc., are expected to be quite different from that in Landau Fermi liquids. A possible statistical mechanical model for the quantum critical fluctuations in diverse systems with such properties is the 2D dissipative quantum XY model. Exact relations between the density correlations and singular irreducible vertices due to coupling of Fermions to the topological excitations of the 2D dissipative quantum XY model are used to derive results which are measurable only recently due to advances in experimental techniques. The density correlations are unusual at all momenta ${\bf q}$ and energy $\om$, from the hydrodynamic limit to that for large momenta and energy. The hydrodynamic limit together with the continuity equation gives the linear in T resistivity.
The density correlations are almost independent of frequency up to a high frequency cut-off for $q_{ZB} \gtrsim q >> \om/v_F$; $q_{ZB}$ is the Brillouin zone boundary and $v_F$ is the Fermi-velocity. The results should be applicable to loop-current quantum criticality in cuprates, and to 2D Fe based compounds near their antiferromagnetic quantum-criticality. The relation of the results to the temperature and frequency dependent conductivity and to Raman response is also discussed. 
\end{abstract}
\pacs{}
\date\today
\maketitle

\section{Introduction}
The dynamic structure function $S(\fq,\om)$ of a system is related to the absorptive part of the density correlation function $\chi_{\rho \rho}(\fq,\om)$ through use of linear response theory and detailed balance \cite{Pines-Nozieres}:
\be
\chi"_{\rho \rho}(\fq,\om) = -\pi \big(1-e^{-\beta \om}\big)S(\fq,\om).
\ee
It contains information on the single-particle excitations, the incoherent multi-particle excitations as well as the collective oscillations. $\chi"_{\rho \rho}(\fq,\om) $ is related to the current-current correlations at long wave-lengths through the continuity equation. It enshrines  the f-sum sum-rule and the sum-rule imposed by particle conservation. It has played a historic role in strongly interacting condensed matter physics. Feynman \cite{superfluid-Feynman1953,  superfluid-Feynman1954} used the sum rules on $S(\fq,\om)$ to derive that in superfluid Helium the only long wave-length excitations are sound waves as well as to derive rotons at higher momenta, both suggested earlier by Landau \cite{superfluid-Landau1941} through phenomenological arguments. Landau theory of Fermi liquids \cite{FermiLiquid-Landau1956, FermiLiquid-Landau1957} has provided precise relations at long wave-length and low energies for neutral systems as well as the collective longitudinal (plasmons) and transverse excitations in charged systems. The developments of the field theory methods to statistical physics \cite{AGD, Nozieres-book} has provided detailed understanding of how Pauli principle restricts the form of the multi-particle excitations in interacting Fermi-liquids at both low and high energy and momenta.

This paper concerns the density correlation function in some metals \cite{Ginzburg-rev, CMV_Lorentz, HvLRMP2007, Analytis, ThemopowerFe, Carrington2013}, which do not obey the quasi-particle paradigm of Fermi liquids in  a characteristic region in the vicinity of a quantum critical point. This class includes the cuprates \cite{Ginzburg-rev}, the Fe-based compounds \cite{Analytis, ThemopowerFe, Carrington2013}, and the heavy-Fermion compounds \cite{HvLRMP2007}. In order to understand a wide variety of anomalous properties - the temperature dependence of the resistivity, the frequency dependence of the optical conductivity, the Raman response, the specific heat or thermopower, as well as the anomalous nuclear relaxation rate,  through a single phenomenological hypothesis, it was suggested that there must exist fluctuations of some operator, to which the fermions couple, which over  over most of the momentum $\fq$ has a form \cite{CMV-MFL}: 
\be
\label{chi}
{C}"(\fq,\om) &=& -N(0) \frac{\om}{T},~~ \om<< T, \\ \nonumber
&=& -N(0) , ~~ T << \om << \om_c, \\ \nonumber
&=& 0,~~ \om >> \om_c.
\ee
$N(0)$ is the order of the density of states. Since the real part ${C}'(\fq,\om)$ has a $\log (T/\om_c)$  singularity for $T >> \om$, fluctuations around a quantum-critical point was indicated. It was not clear, when this idea was proposed, what the operator is  whose fluctuations are of the form (\ref{chi}). 

Scattering of fermions from fluctuations of the form (\ref{chi}) gives a single-particle self-energy which is linear in $max(\om, T)$ and nearly independent of momentum \cite{CMV-MFL, CMV-IOP-QCF} for $\om \lesssim \om_c$ and constant thereafter. This prediction required technical developments in ARPES to be verified fully through  the measurement of the inelastic part of the single-particle spectral function \cite{VallaPRL2000, KaminskiPR2005, Zhu-CMV-MFL}. The resistivity (and the optical conductivity) was obtained earlier \cite{CMV-MFL} from a simple microscopic calculation using the renormalized one-particle propagators which include such self-energies.  The calculation works only due to the astonishing assumption of q-independence (or very weak q-dependence) of the fluctuations of (\ref{chi}) throughout the Brillouin zone. Then there are no vertex corrections in the calculation of the conductivity.  No alternative form for fluctuations nor calculations based on any other ideas have given such results. Similarly the Raman response, on which a brief comment is made later, is also obtained. The same fluctuation spectra has been deduced as the glueing interaction for d-wave fermions in a family of cuprates through analysis of high resolution ARPES experiments in the superconducting state \cite{BokScienceADV}.
 
 From the requirement that the density fluctuations $\chi"_{sc}$ must be  proportional to $q^2$ in the hydrodynamic regime \cite{Kotliar-epl1991}, it was suggested that they are given for $v_Fq << \om$ by,
\be
\label{chihydro}
\chi"_{sc}(\fq,\om) &=& -\kappa \frac{xq^2}{\om(\om^2 +\pi^2 \lambda^2x^2)}, \\ \nonumber
&x& = max(|\om|,\pi T).
\ee
Here $\chi_{sc}$ is the density fluctuations of "screened" fermions or fermions with short-range interactions. The relation of the screened density response and the actual density response in a charged system of Fermions is well known \cite{Pines-Nozieres} and given below.  Here $\kappa$ may be identified to be the compressibility and $\lambda$ is a coupling constant. The frequency and temperature dependence of the long wavelength density correlations is unlike a Fermi liquid in the hydrodynamic regime, in that the effective diffusion constant in (\ref{chihydro}) is frequency and temperature dependent \cite{Shekhter-V-Hydro}. Eq. (\ref{chihydro}) was designed to give the observed temperature dependence of the resistivity on using the continuity relation between the current-current and the screened density correlations, as well as the frequency dependence of Raman scattering. As mentioned above, such a resistivity was already calculated by scattering of fermions by fluctuations of the form (\ref{chi}).

The singular correlations in cuprates (and in the Fe-based compounds) in the quantum-critical region are obviously not of the density. The quantum-critical fluctuations in the Fe-based compound are due to an anti-ferromagnetic critical point. In hole-doped cuprates, the critical fluctuations are of a translation preserving orbital current order \cite{cmv1997, simon-cmv, Bourges-rev} in the form of multiple loops within a unit-cell. Both models map to the DQXY model in 2D \cite{Aji-V-qcf1, CMV-IOP-QCF}. This model, just as the classical 2D XY model, does not fall in the universality class of the Ginzburg-Landau-Wilson type. The critical correlations are determined by topological excitations in space and time, unlike the soft mode theories of the dynamics of critical phenomena of the Ginzburg-Landau-Wilson type. They have been derived analytically \cite{Hou-CMV2016} and tested by quantum Monte-Carlo calculations \cite{ZhuChenCMV2015, Zhu-2016}. 

The  density correlation function Eq.~(\ref{chihydro}) for $v_Fq << \om$ has been derived \cite{Shekhter-V-Hydro} to follow from the critical fluctuations of the DQXY model of the form Eq.~(\ref{chi}). Our purpose here is to derive the density fluctuations using Eq.~(\ref{chi}) in the other limit and to show that they bear resemblance in their frequency dependence to Eq.(\ref{chi}) itself in the opposite limit. The density-correlations in the entire (${\bf q},\om)$ region are thereby shown to inherit the singularities related to those of the critical correlations. 

The impetus for presenting the results for $S({\bf q}, \om)$ for singular Fermi-liquids comes from the recent development of experimental tools \cite{Kogar2014, Kogar2016} through which it can be reliably measured over the entire Brillouin zone and energies up to several eV with high resolution. Such measurements in the past \cite{Roth2014} have been available either through electron energy loss experiments in the forward direction, which only measures plasmons resonances, or through inelastic X-ray scattering \cite{Ghiringhelli2011, LeTacon2011} which gives results with poor energy and momentum resolution in the range of interest for conduction electron density correlations. The calculations lay a basis with which the forthcoming experimental results may be compared. Some preliminary results have been presented \cite{Mitrano_baps}. Some other related matters are also discussed.

\section{Density correlations}

\begin{figure}
 \includegraphics[width=0.8\columnwidth]{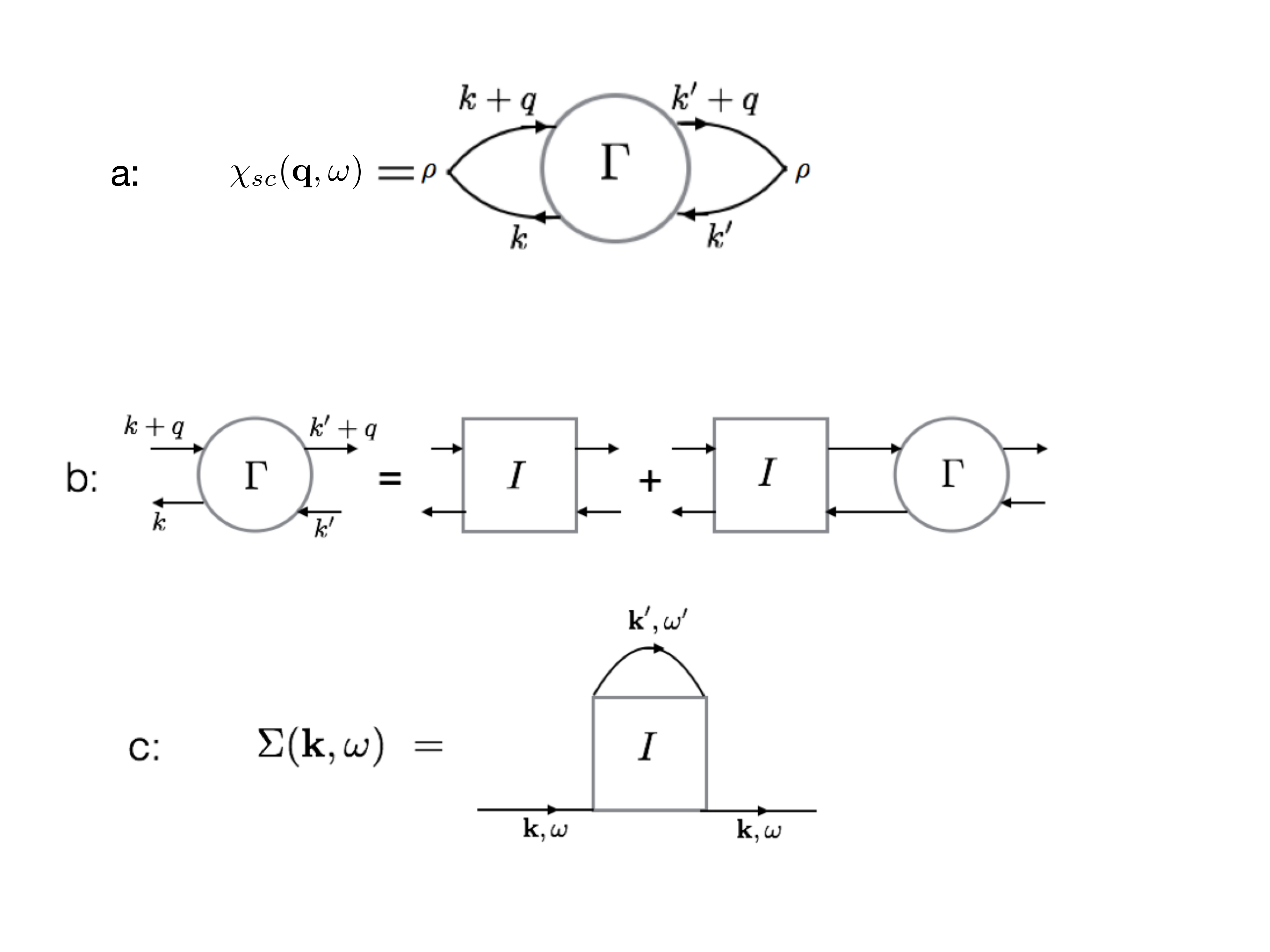}
\caption{(a): Theexact  relation between the density-density correlation function $\chi_{sc}({\bf q}, \om)$ in terms of the total vertex $\Gamma(k,k';q)$ in the particle-hole channel and the single-particle Green's functions. (b): Bethe-Salpeter Equation for the total particle-hole vertex $\Gamma$ in terms of the irreducible particle-hole vertex $I$. The irreducible vertex is defined so that it cannot be cut-into two parts by cutting a particle and a hole line in the channel carrying the energy-momenta $(\omega-\omega', {\bf k-k}')$ (c): The relation of the single-particle self-energy to the irreducible vertex  \cite{Nozieres-book}, with momenta-energy $({\bf q},\om) = 0$. The irreducible vertex is assumed to be regular in the channel carrying $q$ but may have singularities in the other particle-hole channel.}
\label{Fig-Diagrams}
\end{figure}

The screened density correlation function  $\chi_{sc}(\fq,\om)$ is obtained from the total vertex function 
 in the particle-hole channel $\Gamma(k,k',q)$ and the single-particle Green's function $G(k)$, as shown in Fig.(\ref{Fig-Diagrams}-(a)) \cite{Nozieres-book}. (A lower case $k$ stands for momenta and energy. Spin labels will not be explicitly written down in this paper.)  The total vertex is related to the irreducible particle-hole vertex $I(k,k',q)$ by the Bethe-Salpeter equation which is also shown in Fig. (\ref{Fig-Diagrams}-(b)). In a Landau Fermi-liquid, $I(k,k',q)$ is regular. One of the ways in which a Fermi-liquid may be singular is for such vertices to be singular in one of the two particle-hole channels \cite{CMV_Lorentz}. The singular channel is transverse to the one which carries the density, i.e. in the diagram as drawn, it is in the vertical direction in the irreducible particle-hole vertex $I$ as well as the total vertex $\Gamma$ which carries $(k-k')$. 
 
In a charged system of fermions, it is convenient to first consider a screened problem with only short-range interactions in the irreducible vertices and obtain a "screened" $\chi_{sc}(\fq,\omega)$ with the above procedure. The actual density correlations  $\chi_{\rho \rho}(\fq,\om)$ is then obtained from $\chi_{sc}(\fq,\om)$ by summing the polarization graphs \cite{Pines-Nozieres}.
 \be
 \label{chi"}
 \chi"_{\rho \rho}(\fq,\om) = \frac{\chi"_{sc}(\fq,\om)}{|\epsilon(\fq,\om)|^2}, \\
 \epsilon(\fq,\om) = 1-V({\bf q})\chi_{sc}(\fq,\om).
 \ee
 It is therefore enough for a theory to provide $\chi_{sc}(\fq,\om)$ provided $V({\bf q})$ is known. For a 3 D system $V(q) = 4\pi e^2/q^2$. This is also true for a 2D system with 3D electromagnetic fields. However, for a layered system in which the fields are 3D but the electronic correlations are 2D, $V({\bf q})$ depends differently on the momentum perpendicular to the layers $q_z$ and on the 2D momentum in the layers ${\bf q}_{2D}$ .
In (\ref{chi"}) we have taken the background dielectric constant to be 1. Fermions on a lattice always have  inter-band scattering. If one is interested in effects only due to intraband scattering and they are in an energy region smaller than most of the inter-band scattering, one can reasonably replace the 1 in the right side of the equation for $ \epsilon(\fq,\om)$ by an appropriate $\epsilon_0$. 

We will be concerned only with $\chi_{sc}(\fq,\om)$ in this paper with $\fq$ in the plane, since for the problems of interest the many body effects come dominantly from scattering within the planes. (The momentum in the plane in the rest of the paper will be denoted by $\fq$.) This will mean that we will not be concerned with plasmons which are expected to dominate the response at small ${\bf q}$ for $\om >> v_Fq$. It is expected that the physics of the plasmons remains essentially unaltered from that in a Fermi-liquid by the singularities of interest. It should be noted that the current-current correlation is directly related to $\chi_{sc}(\fq,\om)$ rather than to $\chi_{\rho\rho}(\fq,\om)$:
\be
\label{chijj}
Lim_{{\bf q} \to 0}\chi"_{jj}(\fq, \om) = Lim_{{\bf q} \to 0} \frac{\om^2}{q^2} \chi"_{sc}(\fq,\om).
\ee
This together with the relation that the real part of the conductivity 
\be
\label{sigma}
\sigma(\om, T) = \frac{1}{\om} \chi"_{jj}(0, \om)
\ee
gives the observed anomalous conductivity.

 \section{Irreducible vertex  and the density correlation function of fermions coupled to the 2D- dissipative Quantum XY model}
 
 As mentioned already, a number of  2D-metals of experimental interest have quantum-phase transitions which may be described by the dissipative quantum XY model (DQXY). These include models with ferromagnetic or antiferromagnetic transitions with appropriate anisotropies and the loop-current transition in cuprates and in some iridates. The critical fluctuations in such models are not in the density channel, but they change the density-density correlations irrespective of their microscopic origin from those of a fermi-liquid in a characteristic manner, which may be measurable. 
 
 Let us denote the propagator of the fluctuations of the 2D-DQXY model by ${\cal C}(\fq,\om)$ and let them couple to fermions
 by a vertex $g({\bf k,k}')$. ${\cal C}(\fq,\om)$ is closely related to ${C}(\fq,\om)$ of Eq. (\ref{chi}) assumed in the phenomeonology. An irreducible vertex $I(k,k';q)$ may be constructed as
   \be
I(k,k';q) = g(k,k'){\cal C}(k-k') g(k',k)
 \ee
 For the quantum phase transition of the DQXY model of interest, ${\cal C}(\fq,\om)$ is found through RG calculations \cite{Hou-CMV2016},
 and checked by quantum Monte-carlo calculations \cite{ZhuChenCMV2015, Zhu-2016},
 \be
\label{chi-tr}
{\cal C}''({\bf k-k}', \om-\om', T) &\approx& - \chi_0  \tanh\left(\frac{(\om-\om')}{\sqrt{(2T)^2 + \xi_{\tau}^{-2}}}\right) \frac{1}{|{\bf k-k}'|^2 + \xi_r^{-2}}
\ee  
$\xi_{\tau}$ and $\xi_r$ are the correlation length in time and in space, respectively. The 2DQXY model has the property that in dimensionless units $\xi_r \propto \ln(\xi_{\tau})$, or a dynamical critical exponent $z \to \infty$.  

These results come from the calculations of the correlations of  topological excitations of the 2DQXY model, which dominate the critical dynamics. Although at the starting level in the pure electronic Hamiltonian they arise from particle-hole excitations with strong interactions (not confined to near the chemical potential), their final form represents composite objects which bear no resemblance to them, in contrast to the collective excitations such as spin-waves which do. There is no choice but to represent their coupling to fermions as an irreducible vertex. 

The effective coupling $g({\bf k,k}')$ of the fermions to the critical fluctuations are through the potential energy of the 2D-rotors or to the kinetic energy of the rotors \cite{CMV-IOP-QCF}. It is important for us here only to note that symmetry ensures that the magnitude of both these couplings at small momenta are proportional to the magnitude of the momentum transfer, $|{\bf k-k}'|$. Therefore,
 \be
I(k,k',q) \propto \frac{g_0^2|{\bf k-k}'|^2}{|{\bf k-k}'|^2 + \xi_r^{-2}}, 
\ee
where $g_0$ is a coupling constant. In all diagrams shown in Fig.(\ref{Fig-Diagrams}), the dependence on ${\bf k}$ and ${\bf k}'$ is integrated over 2D. Since the topological defects have a spatial variation of the order of a lattice constant, there is no important $q$-dependence either. The momentum dependence of the irreducible vertex is then un-important.  We will therefore ignore the ${\bf k, k',q}$ dependence of $I(k,k',q)$ in the calculations.  Given also that the frequency dependence depends on $(\om-\om')$, we can represent 
$I(k,k',q)$ by a wiggly line passing between a particle and a hole line with arbitrary momentum transfer. This makes the calculations straight-forward. 

To construct the vertex $\Gamma$ from the irreducible vertex $I$ using the Bethe-Salpeter equation, we also need the single-particle Green's function $G(\fk,\om)$. The single-particle self-energy $\Sigma(\fq,\om)$ can be evaluated (non-perturbatively) in terms of the irreducible vertex - See Fig.(\ref{Fig-Diagrams}-(c)). (A clear derivation of the relation of the self-energy to the irreducible vertex is given in Ref. (\onlinecite{Nozieres-book})). This gives the marginal fermi-liquid (mfl) form of the self-energy near criticality:
\be
\Sigma(\fq,\om) = \lambda\Big(\om~\log \Big|\frac{\om_c}{x}\Big| - i (\pi/2) ~x \Big); ~ \lambda = \pi g_0^2 N(0)\chi_0.
\ee
Here $x = max(\omega, \pi T)$. This form is only true for a circular fermi-surface with constant fermi-velocity. If the velocity varies around the fermi-surface, the self-energy has a corresponding angular dependence in magnitude but the frequency dependence remains the same \cite{Zhu-CMV-MFL}. This form of self-energy has an upper cut-off $\omega_c$. For $\omega \gg \omega_c$, the self-energy is a constant.

\subsection{Hydrodynamic region}

In Ref.(\onlinecite{Shekhter-V-Hydro}), the vertex $\Gamma$ and the density correlation $\chi({\bf q}, \om)$ have been derived by evaluating the diagrams in Fig.(\ref{Fig-Diagrams}) in the limit $v_Fq << \om$.  The fact that the imaginary part of $\Sigma$ is itself proportional to $\om$ leads to subtle complications due to branch cuts in $G({\bf k}, \om)$ for calculations in this regime, which are dealt with following Eliashberg's method for calculation of sound velocity in liquid He$^3$. The results are that at for $v_Fq << \om$ and $T << \om$,
\be
\label{chi-hydro}
\chi"_{sc}({\bf q}, \om) &\approx &  N(0) (v_F^2/2)\frac{q^2}{(\om^2 +|\overline{\Sigma}(\om)|^2)}.
\ee
In Ref. (\onlinecite{Shekhter-V-Hydro}), residual Fermi-liquid factor corrections in (\ref{chi-hydro}) have also been given which replace $N(0)$ by the Fermi-liquid compressibility. $\overline{\Sigma}(\om)$ is the same as $\Sigma(\om)$ except that its real part has $\log (2\omega_c/\om)$, rather than $\log (\omega_c/\om)$. Eq. (\ref{chi-hydro}) is then the same as Eq. (\ref{chihydro}) except that the upper cut-off is at $\om \sim 2\omega_c$. It properly obeys the continuity equation and the compressibility sum-rule and gives the conductivity as discussed above.  The imaginary part of the self-energy as well as the logarithmic part in its real part  has $|\om|$ replaced by $\pi T$ for $T >> |\om|$. Then the dc resistivity (in fact resistivity for $\om << T$) is $\propto T$ without any logarithmic correction to the mass. The cancellation of the diverging mass in the zero-frequency limit is due to the fact that the theory obeys the continuity equation or the equivalent Ward identity. 

The hydrodynamic regime in a (pure) marginal Fermi-liquid is worth dwelling on a bit.  A hydrodynamic regime is usually defined for $q \ell <<1$ and $\om \tau <<1$, where $\ell$ is the mean-free path and $\tau$ is the scattering time. Then the density correlations have a diffusive form with the diffusion coefficient $D \propto v_F^2 \ell$. Here we have the scattering rate itself $\propto max(\om,T)$, so that the limit in which the density correlations in a marginal Fermi-liquid are hydrodynamic is $q v_F << \om << T$, when one can write them in a diffusive form $\propto D(T)q^2/(i\om + D(T)q^2)$ with a diffusion coefficient D(T) proportional to the temperature. For $ \om >> T >> v_Fq$, we may need a new term for the derived behavior, also observed through optical conductivity experiments - I suggest "quantum-hydrodynamic" regime.

It is important to note that the upper cut-off in frequency for the density fluctuations is calculated to be $2 |\omega_c|$, whereas it is $|\omega_c|$ in the fundamental fluctuation spectra,  as it is in the one-particle or one-hole spectral function. As derived in \cite{Shekhter-V-Hydro}, this comes about
because the density fluctuations acquire the upper cut-off which is the sum of the extent of the branch-cuts in the one particle and in the one-hole spectra. 

\subsection{High $q$ region}

Now let us consider the limit that $v_Fq >>\om$, where the calculation does not require the subtle mathematical considerations in the complex plane in the opposite limit. In both limits, essential use is made of the effective momentum-independence of the irreducible vertex $I$. This allows a simple evaluation of the T-matrix in
Fig. (\ref{Fig-Diagrams}-(b)) because the summation over momentum can be carried out in any particle-hole section independently of others.  Also, For $T \to 0$, the irreducible propagator is a constant donated by $I(\Omega)$, in the frequency range in the vertical channel $- \omega_c \lesssim \Omega \lesssim \omega_c$. 

\begin{figure}
 \begin{center}
 \includegraphics[width=0.6\columnwidth]{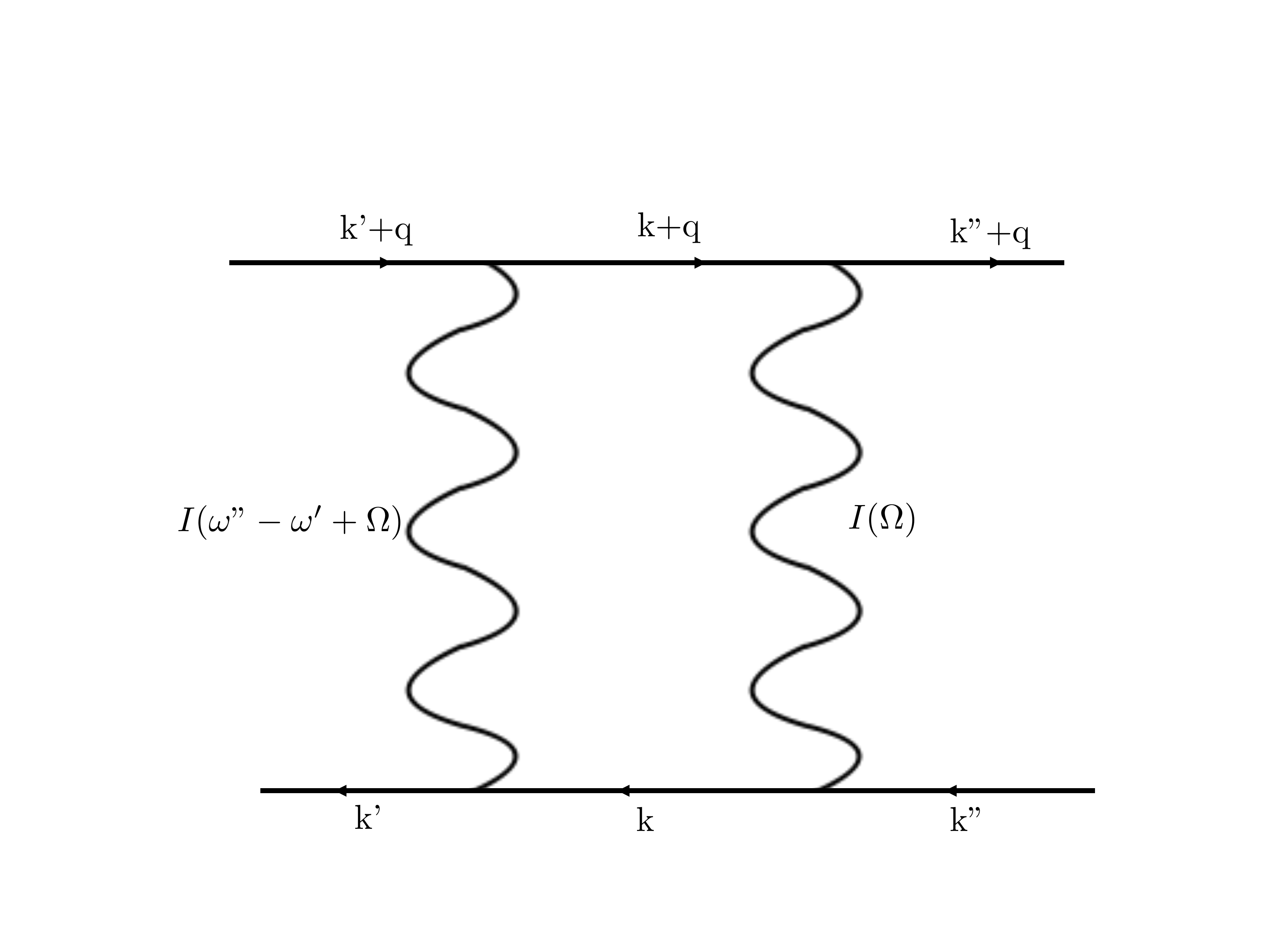}
\end{center}
\caption{Diagram with one particle-hole section of the Bethe-Salpeter equation relating the full vertex to the irreducible vertex, which is calculated in the text. The irreducible vertices depend only on the energy transfer. In such a case, the irreducible vertices may be expressed simply by a propagator connected to the fermions by a point vertex.}
\label{Fig-ph}
\end{figure}

Let us first consider the series for $\Gamma = (I + I GGI + IGGIGGI+...)$, and calculate the second term represented by Fig. (\ref{Fig-ph}). 
 This diagram (\ref{Fig-ph}) gives for external energy-momenta $k' = ({\bf k'}, \om'),k"=({\bf k"}, \om"),q=({\bf q}, \om)$,
\be
2 i (\pi T)^2 \sum_{\Omega_m} I(\Omega_m)  I(\om_n" +\Omega_m - \omega'_l) ~ \times \\ \nonumber
\sum_{\bf k} G({\bf k+q},i(\omega_n"+\Omega_m + \om)) G({\bf k},i(\omega_n"+\Omega_m)).
\ee 
For $v_Fq >>\om$, the self-energy is unimportant in the $G's$. (On the other hand, for $v_Fq << \om$, the branch cuts introduced by the self-energy are crucial \cite{Shekhter-V-Hydro}.) In doing the sum over the frequency $\Omega_m$, it is important to take into account that $\epsilon_{\bf k}$ have finite upper and lower limits given by the bandwidth $-W$ to $W$. For considering the contribution in the sum due to the poles in $G's$, we note that since $I$'s are constant for their argument in the interval ($-\om_c,\om_c$), only a restriction on $(\om_n" -\om_l')$ in the range $(-\om_c, \om_c)$ is placed on them if we sum over $\Omega_m$ over the same range only for the product of the $G$'s alone for the contribution of their poles . The pole contribution then yields a result of $O(\om/qv_F)$, which can be ignored. The important contribution comes due to the finite band-width of $\epsilon_{\bf k}$. For this part we sum over ${\bf k}$ first:
\be
\int d\theta_{kq}N(0) \int_{-W}^{W}d\epsilon_{\bf k} \big(\frac{1}{ i(f_l + \om) -\epsilon_{\bf k+q}}\big)\big(\frac{1}{if_l-\epsilon_{\bf k}}\big),
\ee
where $f_l = \om_n" + \Omega_m$. On doing the integration over $\epsilon_{\bf k}$ assuming free-electron bands with a finite band-width, one gets
for $\om << v_Fq$,
\be
\int d\theta_{kq} \frac{1}{vq \cos \theta_{kq} + q^2/2m} \ln \Big( \Big|\frac{W + if_l +vq \cos \theta_{kq} +q^2/2m}{-W +if_l +vq \cos \theta_{kq} + q^2/2m|}\Big|\Big|\frac{-W + if_l }{W +if_l }\Big|\Big).
\ee
We note that there is no dependence on $\om$ in the limit considered. Consider the q-dependence. At small $vq/W <<1$, we may expand the logarithm to find that the first term is a constant, the term linear in $v_Fq$ vanishes on doing the angular integral, the term of $O(v_f^2 q^2)$ gives a non-zero contribution. Noting that in general a smooth q-dependence $F(q)$ is obtained and that there is only logarithmic dependence on $f_l$ for general $f_l$ which vary on the scale of $\om_c$ which is comparable to $W$ as known from the ARPES experiments \cite{BokScienceADV, ZhuVPRL2008}. The leading contribution to the sum over $\Omega_m$ in Fig. (\ref{Fig-ph}) may thus be estimated from 
\be
N(0) F(q) i\pi T\sum_{\Omega_m}I(\Omega_m)  I(\om_n" +\Omega_m - \om'_l) 
\ee
The most important result then is that (to logarithmic correction) the significant dependence on frequencies is only through $(\om"-\om')$, i.e. two irreducible ladder produces similar frequency dependence  as the single  irreducible vertex. Consider the result at $T=0$ and convert the sum over the frequency to an integral. The result then is a function of $(\omega'-\omega")$, which comes from the convolution of the integral over $\Omega $ in the limits ($-\om_c, \om_c$). If $IN(0) <<1$, this contribution is small to the same order as well as of $O(\omega'-\omega")/\om_c$. 

If we consider larger number of rungs in the ladder, the magnitude is reduced in each order by $IN(0)$ and the frequency carried by the irreducible vertex is further restricted. Each extra rung brings in an extra fermion loop so that the signs alternate. The net effect is a reduction in the leading vertex and this reduction is analytic and considerable. We conclude therefore that the total vertex is given well by the  irreducible vertex for $v_Fq >> \omega$.

Finally we come to calculating the density-correlation function, i.e. Fig.(\ref{Fig-Diagrams}-(a)). Now we have two pairs of Green's function with independent momentum integration but coupled to each other in frequencies which appear in the vertex function. The strategy 
of the calculation is the same as in evaluating Fig. (\ref{Fig-ph}); we drop the self-energy compared to $v_Fq$ (noting at the end that the external frequency $|\omega|$ has a cut-off at $2|\omega_c$ due to the self-energy, as discussed earlier), and do not consider the contributions of the poles because it is of $O(\om/vq)$. The leading contribution to the imaginary part comes from the imaginary part of $I$ and the real parts
of the products of the pairs of GreenÕs functions. The result is

\be
\chi"_{sc}(\om,q) &=& i \pi T \sum_{\om'_m,\om"_n} I (\om'_m-\om"_n) \times \\ \nonumber
&  & \int d\theta_{k"q} \frac{1}{vq\cos \theta_{k"q} + q^2/2m} \ln \Big( \Big|\frac{W + i\om"_n+vq \cos \theta_{k"q} +q^2/2m}{-W +i\om"_n+vq \cos \theta_{kq} + q^2/2m|}\Big|\Big|\frac{-W + i\om"_n}{W +i\om"_n}\Big|\Big) \times \\ \nonumber
&  &\int d\theta_{k'q} \frac{1}{vq\cos \theta_{k'q} + q^2/2m} \ln \Big( \Big|\frac{W + i\om'_m+vq \cos \theta_{kq} +q^2/2m}{-W +i\om'_m+vq \cos \theta_{kq} + q^2/2m|}\Big|\Big|\frac{-W + i\om'_m}{W +i\om'_m}\Big|\Big).
\ee
The most important result is that for $v_Fq >> \om$, there is no dependence of $\chi"_{sc}(\om,q)$ on $\om$ except for a cut-off. Next we consider the q-dependence. In the limit that $v_fq << W$ and for $\om =0$ , we may again expand the logarithms and find that the leading term is a constant in $q$, followed by a term proportional to $q^2$; the term linear in $q$ is absent. The actual $q$- dependence depends on the details of the band-structure. We note that there is no restriction put on $\omega$ except that it is smaller than $v_Fq$, which in effect
amounts to $\omega$ less than $O(W)$. In cuprates, this is similar to $2\omega_c$. Also the results in the crossover region must be continuous with the results for $\omega \gtrsim v_F q$, where the upper cut-off for $\omega$ is $2\omega_c$. We may write the final result for $v_Fq >> \om$ and at $T=0$ as
 
\be
\label{chisclargeq}
\chi"_{sc}({\bf q}, \om) \approx - sgn(\om) G(q) \lambda N(0) I_0, ~ -2\om_c \lesssim \om \lesssim 2\omega_c, G(q) =(G_0 + G_2 q^2 a^2 + ...)
\ee
The real part, $\chi'_{sc}({\bf q}, \om)$, using Kramers-Kronig transformation is $\propto \log |2\omega_c/\om|$.
At finite $T$, $sgn(\om)$ in (\ref{chisclargeq}) is expected to be changed to $\tanh(\om/2T)$.

The results above are for $T=0$ and at criticality, i.e. for $\xi_{r}^{-1} = \xi_{\tau}^{-1} =0$. For finite $T$ and at criticality, they generalize to Eq. (\ref{chi}) easily. For departure from criticality on the Fermi-liquid side, the low energy (and long wavelength) properties revert to Fermi-liquid but only at low temperatures and energies compared to the cross-ver scale. For much higher temperatures and energies, there is a crossover back to the same results as at criticality.

The calculation above is straightforward. The leading results, ignoring logarithmic corrections, etc., are very simple. The calculation itself
is done in unusual limits compared to Fermi-liquid calculations, because of the unusual nature of the irreducible vertex and its large cut-off. Calculations of the regime between the quantum-hydrodynamic and the opposite limit are forbidding and so therefore is the determination of the cross-over between the two.

Let us discuss the compressibility sum-rule,
\be
\int_{0}^{\infty}d\om ~\om ~
S(q,\om) = q^2/2m,
\ee
or its genralization for a tight-binding band \cite{Tremblay-sumrule}.
In the hydrodynamic region(s),  since $S(q,\om) \propto q^2/\om^2$, this is satisfied to logarithmic order and an appropriate cut-off function is required to take care of the logs. In the collisionless limit, it is satisfied with the constant term $G_0$ in $G(q)$ only if the upper limit for this regime is fixed at $w = v_Fq$. For higher powers of $q^2$ in $G(q)$, it is satisfied if the range of integration is not decided by $q$ but is of $O(\omega_c)$.
As discussed, the crossover between the two regions is very hard to calculate.

\section{$S({\bf q}, \om)$ for AFM quantum-criticality of the XY class}

The transport anomalies as well as the entropy in the AFM - quantum critical fluctuation region of several compounds are similar to that in the cuprates. It has been argued that due to either Ising anisotropy with incommensurate AFM fluctuations or due to XY anisotropy itself, the AFM quantum critical fluctuations in 2D also belong to the class of the dissipative quantum XY model. The crossover to the anisotropic quantum critical point \cite{V_crossover} is expected to be over a much wider region of temperature above the $T=0$ quantum critical point than the range of $(T-T_c)/T_c$ near classical critical points. It is of interest to ask what will be the dynamic structure function $S({\bf q}, \om)$ in this case. 

The quantum-critical fluctuations in the 2D- DQXY model for AFMs are expected to be of the same form as Eq.~(\ref{chi-tr}), except that (i) $({\bf k-k}')$ is replaced by $({\bf k-k' - Q})$, where ${\bf Q}$ are the set of AFM Bragg vectors, and (ii)  ${\bf k}$ and $({\bf k'+Q})$ carry opposite spin. 
The vertex $g({\bf k,k'; Q})$ is also modified so that it is proportional to $({\bf k-k' - Q})$. This is sufficient to give a linear scattering rate nearly independent of momentum and a resistivity due to inelastic scattering $\propto T$ as well as a $T \log T$ contribution to the entropy. Even though  the spin-fluctuations are peaked in momentum near ${\bf Q}$, the self-energy is linear in $(\omega, T)$ all around the Fermi-surface; the entire Fermi-surface is hot. These conditions also imply that the dynamic structure function will also be of the forms given in this paper. These are unexpected predictions for AFM quantum-criticality. If they are satisfied a quite different picture for AFM- quantum criticality in 2D emerges.

\section{Summary}

I summarize here the results for the screened density correlations $\chi"_{sc}({\bf q}, \om)$ for the 2D-DQXY model in relation to several experiments including some forthcoming experiments. Some of these results were proposed for cuprates long ago on the basis of a phenemenological spectra when it was not clear what the critical modes are, nor the relevant statistical mechanical model for them.  It has been argued elsewhere that the 2D-DQXY model is the appropriate model for the quantum-critical fluctuations of the loop-current order in cuprates as well as for incommensurate antiferrromagnetic criticality of 2D metals. The critical correlations of this model are driven by topological defects and are quite unlike the anharmonic soft mode fluctuations which are the province of the Ginzburg-Landau-Wilson form of criticality.  The quantum-hydrodynamic limit of the density correlations which are inherited from such critical fluctuations were calculated earlier; the results for the momentum and frequency region outside such a regime are given in this paper here. Experiments in metals measure
$\chi"_{\rho\rho}({\bf q}, \om)$. The relation between the two correlation functions is given by Eq. (\ref{chi}). Outside the small momentum regime in which plasmons are observed, the frequency dependence of the two is expected to be essentially identical, because outside this region the dielectric function $\epsilon({\bf q}, \om)$ is frequency independent up to the high frequency cut-off. 

The observed linear temperature dependence of the resistivity is consistent with the derived $\chi"_{sc}({\bf q}, \om)$ through Eqs. (\ref{chijj}) and (\ref{sigma}). This tests the hydrodynamic limit of the correlation function. No alternative explanation for the linear resistivity exists.
The frequency dependent conductivity and the Raman response test the results for ${\bf q} \to 0$ over a range of frequency up to about 0.5 eV. In comparing the measured conductivity, it is important to bear in mind that there is an upper cut-off in the fluctuation spectra (or of the irreducible vertex) such that the real part of the self-energy modifies the optical response from the scale-invariant response which is a continuation of the zero-frequency response at above about 0.2 eV. The gradual departure from the scale-invariant response depends on the sharpness of the cut-off. This is important to emphasize as it has led to some confusing fits to frequency dependence of the conductivity in an {\it intermediate scale} to power-law forms. Such fits to a problem born of infra-red singularities as at quantum-critical points do not have significance.
In fact, one should in general expect non-universal features in experimental results at frequencies due to other reasons as well. It is the simplicity of the band-structure of the cuprates that universal features are observed to a good approximation from the lowest frequencies to those up to about 0.2 eV in some measurements and even higher in some other measurements. We should not expect this in the Fe-based compounds due to their considerably more complicated band-structure.

No tools have existed to reliably measure the density fluctuation spectra over the complete relevant momentum and frequency region until recently. Outside the quantum-hydrodynamic regime, the screened density correlations are independent of frequency in this region up to the zone-boundary except for the high frequency cut-off which occurs at about twice the cut-off $\omega_c$ in the irreducible vertex. They will in general have a momentum dependence. The irreducible vertex is also given in experiments such as angle-resolved photoemission which measure the single-particle relaxation rate as a function of frequency and is about 0.5 eV. We can combine the results for all momenta and frequency so that the theory can be easily compared with the experiments. The results in the limit of validity of Eq. (\ref{chi-hydro}) and Eq. (\ref{chisclargeq}) may be written as
\be
\label{collect}
Im~\chi_{sc}({\bf q},\om) &\approx & - Sgn(\om) ~\chi_0\frac{v_F^2 q^2}{\omega^2}, ~ for ~ v_Fq << \omega;\\
& \approx   & - Sgn(\om)~ (G(q)/G_0)  \chi_0 ~ ,for ~ 2\om_c \gtrsim  \om >> v_Fq;\\
& \approx   & ~0 ~ ,for ~\om >> 2\om_c.
\ee
We may conveniently interpolate between the region above and below the cut-off in frequency by requiring that the correlation function fall off as $1/\om^2$ as for free-fermions at high frequency.
\be
\label{interpol}
Im~\chi_{sc}({\bf q},\om) = -\chi_0 (G(q)/G_0) \tanh [(\om_c(q)/\om)^2]
\ee
 with $\om_c(q) = v_F q$ for $v_Fq <<  \om$ and $\om_c(q) = \om_0$ for $2 \omega_c>>\om >>v_Fq$. 
 It is noteworthy that the cut-off in the density correlations at about twice those deduced to be dominant in giving linear in $\om$ scattering rate in the single-particle spectra is expected.  
 
 Finally, it is worth recalling a very interesting but idiosyncratically written paper on the Raman response \cite{Shekhter2010}. A remarkable property of the experiments on Raman scattering in cuprates, which has not been much commented on, is that the results are to a first approximation the same in the $A_{1g}, B_{1g}, B_{2g}$ channels \cite{Dever-Hackl2007}. It should be recalled that the Raman response in periodic solids (measured resonantly) is a current-current correlation in the long wave-length limit in different irreducible representations. The fact that similar unuusual results are found in different channels was understood by Shekhter \cite{Shekhter2010} by a calculation in which the critical fluctuations are used as irreducible vertices in the cross-channel, just as in the calculations for density correlations at small $q$ and their extension presented here.

{\it Acknowledgements}: I wish to thank Peter Abbamonte, Ali Husain and Matteo Mitrano for discussions on the experiments, Elihu Abrahams and Arkady Shekhter on various theoretical issues, and especially, Andrey Chubukov on aspects of the calculations.


\end{document}